\documentclass[pra,showpacs,twocolumn]{revtex4}
\usepackage{dcolumn}
\usepackage{graphicx}
\newcommand{\vcc}[1]{{\vec#1}}
\newcommand{\opr}[1]{\mathaccent 94 #1}
\newcommand{\ket}[1]{\left|#1\right\rangle}

\newcommand{\avr}[1]{\left\langle #1 \right\rangle}
\begin{document}
\sloppy
\title{Rotated and scaled center of mass tomography for several particles}
\author{A. S. Arkhipov$^1$}
\email{antoncom@id.ru}
\author{Yu. E. Lozovik$^1$}
\email{lozovik@isan.troitsk.ru}
\author{V. I. Man'ko$^2$}
\email{manko@sci.lebedev.ru}
\affiliation{$^1$ Institute of Spectroscopy RAS, 142190 Troitsk, Moscow region, Russia}
\affiliation{$^2$ Lebedev Physical Institute, Leninsky Prospect 53, 117924 Moscow, Russia}
\date{\today}
\begin{abstract}
The tomographic map of quantum state of a system with several degrees of freedom which depends on
one random variable, analogous to center of mass considered in rotated and scaled reference frame
in the phase space, is constructed. Time evolution equation of the tomogram for this map is given
in explicit form. Properties of the map like the transition probabilities between the different
states and relation to the star product formalism are elucidated. Example of multimode oscillator
is considered in details. Symmetry properties in respect to identical particles permutations are
discussed in the framework of proposed tomography scheme.
\end{abstract}
\maketitle

\section{Introduction}\label{Introduction}
Recently \cite{Tom96} a new formulation of quantum mechanics was suggested. This new formulation
uses {\it non-negatively defined} probability distribution function to describe quantum states,
called marginal distribution \cite{BerBer,KV}, or tomogram. This function can be considered as an
analog of known quasidistribution functions like nonnegative Husimi Q-function \cite{Hus40} or
Sudarshan-Glauber P-function \cite{Sud63,Gla63}. Tomographic approach was initially developed for
one-mode systems, in this case the quantum state is described by the density matrix $\rho(q',q'')$
\cite{Land27,vonNeum32} or by symplectic tomogram $w(X,\mu,\nu)$ \cite{Mancini95}. Here density
matrix is the function of $2$ variables and tomogram is the function of $3$ variables. Seeming
overcompleteness of the tomographic description is balanced by the fact that quantum tomogram is
a homogeneous function \cite{Wunsche,RosaPRA}.

The state of the system with $N$ degrees of freedom is described by density matrix
$\rho(\vcc q',\vcc q'')$, the function of $2N$ variables. Then, what is the generalization of
quantum tomogram, will it depend on $3N$ variables or on $2N+1$ variables? Such generalization
was developed for the tomogram, depending on $3N$ variables (usual symplectic tomography)
\cite{Ariano96,MarmoMMJPhys2002}. In this paper we propose the tomographic map with only one random variable
({\it i.e.} $2N+1$ variables totally) and discuss its properties in details.

Quantum tomography becomes popular nowadays for a number of reasons. It was developed not only for
continuous variables like position but also for discrete spin variables \cite{DodonPLA,WeigertPRL,
OlgaMJETP,MarmoPhysScripta,Leonhardt,KlimovJPA,CastanosJPAJOB}. First, tomographic representation
operates with the values, which are directly measured in experiments, for example non-classical and
coherent states of light \cite{Raymer1993,Ariano2003,Mlynek,WVogelReview,RaymerWFHTPRA1993,
RaymerMcAlisterPRL,Raymer2D1996,Raymer2D2000} or matter optics \cite{Risley1990,Mlynek1991,Pritchard1991,Mukamel1995}
experiments. Second, the tomogram is non-negative, and this must attract the attention of those who
deal with computer simulations of quantum systems \cite{LSA}. Many problems in this field, {\it e.g.}
for Path Integral Monte Carlo simulations, result from the use of the alternating-sign, or even
complex, values to describe the quantum state (for example, sign problem in Fermi-systems simulations).
The latter leads to difficulties with convergence which can be overcome by quantum tomography
approach. These facts along with the tomography applications in quantum computations and entanglement
(see, {\it e.g.}, \cite{AndManEPR}), as well as in the theory of information and signal analysis
\cite{RaymerMcAlisterPRL}, show that the development of convenient and simple tomographic map for
the case of many particles (or, which is the same, many modes) is a task of great significance.

The paper is organized as follows. In Sec.~\ref{TGdefEvol} we present the definition of
tomographic scheme with one random variable, elucidate some of its useful properties and discuss
the physical meaning of the map. In Sec.~\ref{States} we derive the equations describing quantum
evolution, stationary states, quantum transitions and rules for average values calculation for the
proposed tomography map. Some examples of state description using the developed approach are given
in Sec.~\ref{Examples} and symmetry of the map with respect to particles permutations is discussed
in Sec.~\ref{Permutations}. The work is summarized in Sec.~\ref{Conclusion}.

\section{One random variable tomography}\label{TGdefEvol}
\subsection{Definition of the tomographic map}
We begin with the one-dimensional (1D) case of a particle with continuous degree of freedom (in
this paper we do not consider spin variables, but generalization of the formalism is
straightforward). Quantum mechanics states that we know 'everything' about the system if we know
density matrix. In practice, to obtain any information about the system we have to measure some
quantities, for example coordinate $q$ or momentum $p$. It is also possible sometimes to measure
an intermediate quantity, $\mu q+\nu p$, where $\mu,\nu$ are real parameters. Formally, this
quantity (denote it $X$) is {\it coordinate, measured in scaled and rotated reference frame in
the phase space}. It turns out, that the distribution function of $X$ ($w(X,\mu,\nu)$), measured
for all sets of $\mu,\nu$ gives {\it complete quantum mechanical description} of the system, in
the sense that there is a unique correspondence between $w(X,\mu,\nu)$ and density matrix
(see, {\it e.g}, \cite{Tom96,Mancini95,MarmoMMJPhys2002,QfieldTG1998,MendesPhysD}). Note that distribution
function $w(X,\mu,\nu)$ is equal to $\avr{\delta(X-\mu\opr q-\nu\opr p)}$, where $\avr{...}$ is
quantum mechanical average. Then there is, in principle, a possibility of complete experimental
density matrix determination through the set of coordinate measurements.

When we deal with more than one particle and dimension we can consider individual $X_j=\mu_jq_j+
\nu_jp_j$ for every $j$-th degree of freedom. This results in the symplectic tomography representation
\cite{MarmoMMJPhys2002,QfieldTG1998}. Here we are to show that it is enough to work with only one
$X=\sum_jX_j$. To do this, let us consider the generalization of $w(X,\mu,\nu) = \avr{\delta(X-
\mu\opr q-\nu\opr p)}$, where $q,p$ and $\mu,\nu$ becomes the vectors, their products become
scalar products of vectors, while $X$ remains a real number:
\begin{equation}
w(X,\vcc\mu,\vcc\nu) = \avr{\delta(X-\vcc\mu\opr\vcc q-\vcc\nu\opr\vcc p)}\label{QTdefSP}
\end{equation}
Related problems were discussed in \cite{Raymer2D1996,Raymer2D2000}.

Throughout the paper designations are the following. We consider the system of $N$ particles in
$d$ dimensions, the number of degrees of freedom is $Nd$. Vectors are written as $\vcc a$, we use
everywhere the vectors with $Nd$ components, if the otherwise is not mentioned. Designation
$\vcc e$ is used for the vector with all components equal to $1$ ($e_j = 1$). Scalar product of
vectors is designated $a = \vcc b\vcc c$ (meaning $a = \sum_j{b_jc_j}$),
$\vcc a = \vcc b\circ\vcc c$ denotes the componentwise product of vectors ($a_j = b_jc_j$). The
tomogram for usual symplectic scheme is designated as $w_s(\vcc X,\vcc \mu,\vcc \nu)$
($\vcc X,\vcc \mu$ and $\vcc \nu$ with $Nd$ components each); the tomogram with one random
variable is written as $w(X,\vcc \mu,\vcc \nu)$. We also use Planck constant $\hbar = 1$ everywhere.

We can begin the construction of one-random-variable tomography representation from the known star product expressions. In the
framework of star product formalism {\cite{SJETP1957,GBPRA,CFZPRD,BrifMann} every operator is replaced by the function ('symbol'),
depending on specific set of parameters ($y$), products of operators turn into 'star products' of corresponding symbols (star product
problem was discussed also in Ref.\cite{BFFLS}). As a result, one deals with functions only, avoiding operators. For example, using a
pair of operators $\opr D(y), \opr U(y)$, we construct the connection between the symbols $f_A(y)$ and operators $\opr A$:
\begin{eqnarray}
&&f_A(y) = Tr(\opr A\opr U(y)),\\
&&\opr A = \int f_A(y)\opr D(y)dy,\\
&&\int Tr(\opr D(y)\opr U(y)) dy = 1
\end{eqnarray}
For $y=\{X,\vcc\mu,\vcc\nu\}$ one can choose
\begin{eqnarray}
&&\opr U(y) = \delta\left(X-\vcc\mu\opr\vcc q-\vcc\nu\opr\vcc p\right),\label{USP}\\
&&\opr D(y) = (2\pi)^{-Nd}exp\left[i\left(X-\vcc\mu\opr\vcc q-\vcc\nu\opr\vcc p\right)\right],\label{DSP}
\end{eqnarray}
which defines the symbols (denote them $w_A(X,\vcc\mu,\vcc\nu)$) and star product:
\begin{equation}
(w_A*w_B)(y)=\int w_A(y'')w_B(y')K(y'',y',y)dy''dy'
\end{equation}
The kernel of star product $K(y'',y',y)$ is expressed as follows:
\begin{eqnarray}
&&K(y'',y',y)=Tr[\opr D(y'')\opr D(y')\opr U(y)]=\nonumber\\
&&\int e^{-i(kX-X'-X'')}\delta(\vcc\mu''+\vcc\mu'-k\vcc\mu)
\delta(\vcc\nu''+\vcc\nu'-k\vcc\nu)\times\nonumber\\
&&e^{-i(\vcc\mu''\vcc\nu'-k(\vcc\mu''\vcc\nu'+\vcc\mu'\vcc\nu)+(\vcc\mu'\vcc\nu'+
\vcc\mu''\vcc\nu''+k^2\vcc\mu\vcc\nu)/2)}\frac{dk}{(2\pi)^{Nd+1}}\label{w2Kernel}
\end{eqnarray}

For any operator $\opr A$ we have $\avr A = Tr(\opr\rho\opr A)$, therefore $w_{\rho}$-symbol of density
operator $\opr\rho$ is the same as $w(X,\vcc\mu,\vcc\nu)$ defined by Eq.(\ref{QTdefSP}). Density
matrix in any representation is just a matrix element of density operator. Then, finally, we come to
the unique correspondence (invertable map) between the tomogram $w$ and density matrix mentioned above:
\begin{eqnarray}
&&w(X,\vcc\mu,\vcc\nu) = \avr{\delta(X-\vcc\mu\opr\vcc q-\vcc\nu\opr\vcc p)},\label{QTdef}\\
&&\opr\rho = \int w(X,\vcc\mu,\vcc\nu)e^{i(X-\vcc\mu\opr\vcc q-\vcc\nu\opr\vcc p)}
\frac{dXd\vcc\mu d\vcc\nu}{(2\pi)^{Nd}}\label{QTtoDO}
\end{eqnarray}
Density matrix always can be reconstructed from the tomogram $w$ using these equations, so one random
variable tomogram describes quantum state {\it completely}. Note that now the state-describing function
is {\it nonnegative} and depends on $2Nd+1$ variables in contrast to symplectic tomogram, depending on $3Nd$ variables.

\subsection{Properties of $w(X,\vcc\mu,\vcc\nu)$}
It is convenient to consider density matrix in coordinate representation and Wigner formulation of
quantum mechanics \cite{Wigner} to derive the properties and evolution equation for the tomogram $w$.
In the framework of Wigner formalism the state of the system is described by real Wigner function
$W(\vcc q,\vcc p)$ defined in phase space and connected with the density matrix as follows:
\begin{eqnarray}
W(\vcc q,\vcc p) = \int \rho(\vcc q+\frac{\vcc u}{2},\vcc q-\frac{\vcc u}{2})
e^{-i\vcc p\vcc u}\frac {d\vcc u}{(2\pi)^{Nd}}, \label{DMtoWF}\\
\rho(\vcc q',\vcc q'') = \int W(\frac{\vcc q'+\vcc q''}{2},\vcc p)
e^{i\vcc p(\vcc q'-\vcc q'')} d\vcc p \label{WFtoDM}
\end{eqnarray}

Using Eqs.(\ref{QTdef},\ref{QTtoDO}) and Eqs.(\ref{DMtoWF},\ref{WFtoDM}) we obtain:
\begin{eqnarray}
&&w(X,\vcc\mu,\vcc\nu) = \int W(\vcc q,\vcc p) e^{-ik(X-\vcc\mu\vcc q-\vcc\nu\vcc p)}
\frac{dkd\vcc qd\vcc p}{(2\pi)},\label{WFtoQT}\\
&&W(\vcc q,\vcc p) = \int e^{-i(\vcc\mu\vcc q+\vcc\nu\vcc p-X)}w(X,\vcc\mu,\vcc\nu)
\frac{dXd\vcc\mu d\vcc\nu}{(2\pi)^{2Nd}} \label{QTtoWF}
\end{eqnarray}

Usual symplectic tomography map is developed in references \cite{MarmoMMJPhys2002} and \cite{QfieldTG1998}.
The symplectic tomogram $w_s(\vcc X,\vcc\mu,\vcc\nu)$ and Wigner function are connected as follows:
\begin{eqnarray}
w_s(\vcc X,\vcc\mu,\vcc\nu)=\int W(\vcc q,\vcc p)e^{-i\vcc k(\vcc X-
\vcc\mu\circ\vcc q-\vcc\nu\circ\vcc p)}\frac{d\vcc kd\vcc qd\vcc p}{(2\pi)^{Nd}},\label{WFtoQTs}\\
W(\vcc q,\vcc p)=\int e^{-i\vcc e(\vcc\mu\circ\vcc q+\vcc\nu\circ\vcc p-\vcc X)}
w_s(\vcc X,\vcc\mu,\vcc\nu)\frac{d\vcc Xd\vcc\mu d\vcc\nu}{(2\pi)^{2Nd}} \label{QTstoWF}
\end{eqnarray}

Since the Wigner function is connected by invertable maps with both tomograms $w$ and $w_s$ it is
obvious that they contain the same information about the quantum state. In fact one has
\begin{eqnarray}
w(X,\vcc\mu,\vcc\nu) = \int w_s(\vcc Y,\vcc\mu,\vcc\nu)\delta(X - \sum_{j=1}^{Nd}Y_j)d\vcc Y,\label{QTstoQT}\\
w_s(\vcc X,\vcc\mu,\vcc\nu) = \int w(Y,\vcc k\circ\vcc\mu,\vcc k\circ\vcc\nu)
e^{i(Y-\vcc k\vcc X)}d\vcc kdY\label{QTtoQTs}
\end{eqnarray}

The Wigner function is normalized:
\begin{eqnarray}
\int W(\vcc q,\vcc p)d\vcc qd\vcc p = \int \rho(\vcc q+\frac{\vcc u}{2},\vcc q-\frac{\vcc u}{2})
e^{-i\vcc p\vcc u}\frac{d\vcc ud\vcc qd\vcc p}{(2\pi)^{Nd}} = \nonumber\\
\int \rho(\vcc q+\frac{\vcc u}{2},\vcc q-\frac{\vcc u}{2})\delta(\vcc u)d\vcc u d\vcc q =
\int \rho(\vcc q,\vcc q)d\vcc q = 1,\label{WFnorm}
\end{eqnarray}
where we choose the normalization for density matrix $Tr(\opr\rho)=1$. Then the tomogram $w$
is normalized in $X$ variable:
\begin{eqnarray}
\int w(X,\vcc\mu,\vcc\nu)dX = \int W(\vcc q,\vcc p)\delta(k)e^{ik(\vcc\mu\vcc q+
\vcc\nu\vcc p)}dkd\vcc qd\vcc p = 1\label{QTnorm}
\end{eqnarray}

Although the tomogram depends on $2Nd+1$ variables, instead of $2Nd$ for density matrix, the
completeness of physical description is the same for both formulations, due to the fact that
the tomogram is a homogeneous function. Consider the definition (\ref{WFtoQT}) and multiply all
variables in $w$ by a real number $\lambda$:
\begin{eqnarray}
&&w(\lambda X,\lambda\vcc\mu,\lambda\vcc\nu) =
\int W(\vcc q,\vcc p)e^{-i\lambda k(X - \vcc\mu\vcc q - \vcc\nu\vcc p)}
\frac {dk d\vcc q d\vcc p}{(2\pi)} =\nonumber\\
&&\int W(\vcc q,\vcc p) e^{-ik(X - \vcc\mu\vcc q - \vcc\nu\vcc p)}
\frac {dk d\vcc q d\vcc p}{(2\pi|\lambda|)} =
\frac{w(X,\vcc\mu,\vcc\nu)}{|\lambda|},\label{QTLambda}
\end{eqnarray}
where we just made the change of variables $\lambda k\rightarrow k$.

Property (\ref{QTLambda}) make obvious the relations
\begin{equation}
w(X,\vcc\mu,\vcc\nu) = |X|^{-1}w(1,\vcc\mu/X,\vcc\nu/X) \label{QTX}
\end{equation}

For the pure state with wave function $\Psi(\vcc q)$ symplectic tomogram $w_s$ was expressed
in terms of modulus squared of fractional Fourier transform of the wave function \cite{MendesPLA}.
The tomogram $w$ for pure state is given by
\begin{eqnarray}
&&w(X,\vcc\mu,\vcc\nu) = \int d\vcc Y\frac{\delta(X-\sum_{j=1}^{Nd}Y_j)}{(2\pi)^{Nd}
\prod_{j=1}^{Nd}|\nu_j|}\times\nonumber\\
&&\left|\Psi(\vcc q)exp\left[i\left(\vcc q\frac{\vcc Y}{\vcc\nu}-\frac{\vcc q\circ\vcc q}{2}\frac{\vcc Y}
{\vcc\nu}\right)\right]d\vcc q\right|^2 \label{QTPure}
\end{eqnarray}

\subsection{Physical meaning}
We have defined the nonnegative function $w(X,\vcc\mu,\vcc\nu)$ (\ref{QTdef}) completely
describing quantum state. For any set of $\{\vcc\mu,\vcc\nu\}$ it is normalized as a function of $X$,
therefore $w(X,\vcc\mu,\vcc\nu)$ is the set of distribution functions of quantity $X$. Then to
know the quantum state completely one has to consider all sets of $\{\vcc\mu,\vcc\nu\}$ (in practice,
moving with some step) and measure $X=\vcc\mu\vcc q+\vcc\nu\vcc p$ many times for each set: this
yields the distribution function $w(X,\vcc\mu,\vcc\nu)$ for given set of $\{\vcc\mu,\vcc\nu\}$.

Looking at Eq.(\ref{QTX}) we see that we even do not have to know $w(X,\vcc\mu,\vcc\nu)$, the
value of this function in some point in $X$ for all $\{\vcc\mu,\vcc\nu\}$ is enough. This does
not change the scheme of measurements, we still need to measure full distribution function of $X$
for given $\{\vcc\mu,\vcc\nu\}$ (it is necessary to compare the values of distribution function
in different points to be sure that statistical precision is good), but one has to store the smaller
arrays of information.

Property (\ref{QTLambda}) can be used in another way. If $\lambda$ is equal to $\vcc\mu\vcc\mu+
\vcc\nu\vcc\nu$, we can parameterize $\{\vcc\mu,\vcc\nu\}$ by $\lambda$ and $2Nd-1$ angles (to
use the spherical coordinates in the space of $\{\vcc\mu,\vcc\nu\}$). Applying Eq.(\ref{QTLambda})
we come to reduced tomogram with $2Nd$ variables and $\{\vcc\mu,\vcc\nu\}$ located on the sphere
with radius equal unity in $2Nd$-dimensional space. This new tomogram also completely describes
the state and in some cases it can be convenient to use this one in measurements, because it is
easier to sample $2Nd-1$ angle than $2Nd$-dimensional space from $-\infty$ to $\infty$ (see,
{\it e.g.}, \cite{RaymerMcAlisterPRL,Raymer2D2000}). On the other hand such formulation causes
trouble with the derivation of evolution equations and arbitrary average values calculation.

The only remaining unclear point is the meaning of $X=\vcc\mu\vcc q+\vcc\nu\vcc p$. It is the sum
of positions measured in scaled and rotated reference frame in the phase space. But what does it
mean physically? It is impossible to measure $\vcc q$ and $\vcc p$ simultaneously, but sometimes
one can transform $\vcc q$ and $\vcc p$ into the form $\vcc\mu\vcc q+\vcc\nu\vcc p$, for example
mixing the signal beam with local oscillator field (in quantum optics, see \cite{RaymerWFHTPRA1993}
and references therein). Another scheme was proposed in \cite{RaymerMcAlisterPRL}, where $\vcc q$
and $\vcc p$ are mixed due to wave (electromagnetic or matter) propagation through a lens (or an
analog of a lens in atomic optics). Taking into account the present development of science
concerning controlling the Bose-condensates of atoms, this also can be a possible realm of
tomography measurements. Bose-condensate is a coherent macroscopic state of many atoms and it is
described by macroscopic wave function. For example, one can mix two such waves (condensates of the
same atoms), using the first as a signal wave and the second as local oscillator. Varying the
phase difference of the condensates we sample different $\vcc\mu,\vcc\nu$. Probably, the same can
be done in superconductors (where the electrons of superconductivity also form the coherent
macroscopic matter wave), using Josephson junctions.

If we somehow accomplished the scaling and rotation of reference frame in the phase space we can
measure the set of positions in this reference frame $X_j=\mu_jq_j+\nu_jp_j$, but it is enough to
measure their sum, $X=\vcc\mu\vcc q+\vcc\nu\vcc p$. It is analogous to the the position of center
of mass measurement (the sum of coordinates of corresponding vector, to be more precise). Indeed,
the center of mass position is
\begin{equation}
X_{cm}=\sum_jm_jX_j/M=\sum_jm_j(\mu_jq_j+\nu_jp_j)/M,
\end{equation}
where $M=sum_jm_j$ and $m_j$ is the mass corresponding to $j$-th degree of freedom, and $X_{cm}$
can be associated with $X=\vcc\mu\vcc q+\vcc\nu\vcc p$ for some other set of $\{\vcc\mu,\vcc\nu\}$.
We sample all sets of $\{\vcc\mu,\vcc\nu\}$ therefore it is enough to measure the center-of-mass
position in each scaled and rotated reference frame.

Finally, we would like to make the following remark. The storage of arrays representing full
density matrix or tomogram becomes impossible when the number of degrees of freedom growth. If we
use some grid, the number of arrays elements is proportional to $n^{Nd}$, where $n$ is the number
of grid steps. Increasing $Nd$ we soon come to the situation when all data carriers in the world
can not store corresponding arrays. And this is not necessary as the state of the system is
uniquely determined by the one-particle density (through the density functional, see \cite{KohnNL}
and references therein). Then for many-particles systems description we can use reduced density
matrices (one-body, two-body, etc.), and tomography map is constructed for them in the same way
as for full density matrix. Then the situation with reference frame scaling and rotation is
simplified because $\mu$ and $\nu$ are the same for all particles (if one-body density matrix is
considered) and distribution functions are averaged over all particles.

\section{State transformations}\label{States}
\subsection{Evolution equations}
Let us discuss the evolution equation for tomogram $w$. Begin with the most general evolution
equation for density matrix:
\begin{eqnarray}
i\frac{\partial\rho(\vcc q',\vcc q'')}{\partial t} =
\left[\opr H, \rho(\vcc q',\vcc q'')\right]\label{DMEvol}
\end{eqnarray}
Here and throughout the paper we omit the dependence on time $t$, but imply that all functions,
describing the state (density matrix, Wigner function, tomogram) depend on time as parameter. We
consider the Hamiltonian $\opr H = \sum_i{\opr p_i^2/(2m_i)}+V(\vcc q)$. To derive the evolution
equation for tomograms we consider the Moyal evolution equation for Wigner function \cite{OlgaMJETP,Moyal,MarXiv}:
\begin{eqnarray}
\frac{\partial W}{\partial t}+\vcc{\frac p m}\frac{\partial W}{\partial\vcc q}+\nonumber\\
i\left[V(\vcc q+\frac i 2\frac{\partial}{\partial\vcc p})-
V(\vcc q-\frac i 2\frac{\partial}{\partial\vcc p})\right]W=0, \label{WFEvol}
\end{eqnarray}
where $\vcc p/ \vcc m$ means the vector with components $p_i/m_i$ (the equation holds for the case
of different masses for different particles and directions), the operators in the potential $V$
designates the analytical expansion of the potential and use of the products of corresponding
operators. This equation can be easily obtained applying the transform (\ref{DMtoWF}) to the Eq.(\ref{DMEvol}).

To derive the evolution equation for tomogram one applies the transform (\ref{WFtoQT}) to evolution
equation for Wigner function (\ref{WFEvol}). Expanding the potential in Eq.(\ref{WFEvol}), it is
seen that we have to consider the transforms of the following quantities: $\vcc q W$,
$\partial W/\partial\vcc q$, $\vcc p W$ and $\partial W/\partial\vcc p$. The transform
(\ref{WFtoQT}) of $\vcc q W$ is
\begin{eqnarray}
\int \vcc qW(\vcc q,\vcc p) exp[-ik(X - \vcc\mu\vcc q - \vcc\nu\vcc p)]
\frac{dk d\vcc q d\vcc p}{(2\pi)}= \nonumber\\
-i\frac{\partial}{\partial \vcc\mu}\int \frac{W(\vcc q,\vcc p)}{k}
exp[-ik(X - \vcc\mu\vcc q - \vcc\nu\vcc p)]\frac {dk d\vcc q d\vcc p}{(2\pi)}
\label{qWFtoWT2}
\end{eqnarray}
Consider the operator $(\partial/\partial X)^{-1}$, which gives the antiderivative of the
function it works on. Then we have
\begin{equation}
i\frac{e^{-ikX}}{k}=\frac{i}{k}\left(\frac{\partial}{\partial X}\right)^{-1}\frac{\partial}
{\partial X}e^{-ikX}=\left(\frac{\partial}{\partial X}\right)^{-1}e^{-ikX},\label{eikX}
\end{equation}
and Eq.(\ref{qWFtoWT2}) becomes
\begin{eqnarray}
\int \vcc qW(\vcc q,\vcc p) exp[-ik(X - \vcc\mu\vcc q - \vcc\nu\vcc p)] \frac{dk d\vcc q
d\vcc p}{(2\pi)}= \nonumber\\ -\frac{\partial}{\partial \vcc\mu}\left(\frac{\partial}{\partial X}
\right)^{-1}w(X,\vcc\mu,\vcc\nu)\label{qWFtoWT2got}
\end{eqnarray}
Using the same simple operations we obtain the rules of Eq.(\ref{WFEvol}) terms transformation,
which we formally designate as '$\rightarrow$':
\begin{eqnarray}
&&\vcc qW(\vcc q,\vcc p) \rightarrow -\frac{\partial}{\partial \vcc\mu}\left(\frac{\partial}
{\partial X}\right)^{-1}w(X,\vcc\mu,\vcc\nu)\label{1qpWFtoQT}\\
&&\frac{\partial W(\vcc q,\vcc p)}{\partial\vcc q} \rightarrow \vcc\mu\frac{\partial}{\partial X}
w(X,\vcc\mu,\vcc\nu)\label{2qpWFtoQT}\\
&&\vcc p W(\vcc q,\vcc p) \rightarrow -\frac{\partial}{\partial \vcc\nu}\left(\frac{\partial}
{\partial X}\right)^{-1}w(X,\vcc\mu,\vcc\nu)\label{3qpWFtoQT}\\
&&\frac{\partial W(\vcc q,\vcc p)}{\partial\vcc p} \rightarrow \vcc\nu\frac{\partial}{\partial X}
w(X,\vcc\mu,\vcc\nu)\label{4qpWFtoQT}
\end{eqnarray}
Successive application of rules (\ref{1qpWFtoQT}-\ref{4qpWFtoQT}) allows us to transform all
powers of $\vcc q$, $\vcc p$ and corresponding derivatives in Eq.(\ref{WFEvol}). As a result we
obtain the evolution equation for one random variable quantum tomogram $w$:
\begin{eqnarray}
&&\frac{\partial w}{\partial t}-\vcc{\frac\mu m}\frac{\partial w}{\partial\vcc\nu} +
i\left[V\left(-\frac{\partial}{\partial \vcc\mu}\left(\frac{\partial}{\partial X}\right)^{-1}+
\frac i 2\vcc\nu\frac{\partial}{\partial X}\right)\right.-\nonumber\\
&&\left.V\left(-\frac{\partial}{\partial \vcc\mu}\left(\frac{\partial}{\partial X}\right)^{-1}-
\frac i 2\vcc\nu\frac{\partial}{\partial X}\right)\right]w=0 \label{QTEvol}
\end{eqnarray}

\subsection{Stationary states}
For the stationary state with definite energy we can turn from the time-dependent Schr\"odinger
equation (\ref{DMEvol}) to the eigenvalue equation:
\begin{eqnarray}
\opr H\opr\rho_E=\opr\rho_E\opr H=E\opr\rho_E \label{DMEigen}
\end{eqnarray}

Applying the transform (\ref{DMtoWF}) we obtain the following rules of transition from the
equation for density matrix to equation for Wigner function:
\begin{eqnarray}
&&\frac{\partial^2\rho(\vcc q,\vcc q')}{\partial\vcc q^2}\rightarrow(\frac{1}{4}\frac{\partial^2}
{\partial\vcc q^2}-\vcc p^2 + i\vcc p\frac{\partial}{\partial\vcc q})W(\vcc{q,p}),\nonumber\\
&&V(\vcc q)\rho(\vcc q,\vcc q')\rightarrow V(\vcc q+\frac{i}{2}\frac{\partial}
{\partial\vcc p})W(\vcc{q,p})\label{qDqDMtoWF}
\end{eqnarray}
After that, using (\ref{1qpWFtoQT}-\ref{4qpWFtoQT}), we have the eigenvalue equation for the
tomogram $w$ with one random variable:
\begin{eqnarray}
&&\sum_{j=1}^{Nd}\left[\frac{1}{2m_j}\frac{\partial^2}{\partial\nu_j^2}
\left(\frac{\partial}{\partial X}\right)^{-2} -
\frac{1}{8m_j}\mu_j^2\frac{\partial^2}{\partial X^2}\right]w +\nonumber\\
&&ReV\left(\frac i 2\vcc\nu\frac{\partial}{\partial X} - \frac{\partial}{\partial \vcc\mu}
\left(\frac{\partial}{\partial X}\right)^{-1}\right)w=Ew\nonumber\\
&&-\sum_{j=1}^{Nd}\frac{\mu_j}{2m_j}\frac{\partial w}{\partial\nu_j}=\nonumber\\
&&ImV\left(\frac i 2\vcc\nu\frac{\partial}{\partial X} - \frac{\partial}{\partial \vcc\mu}
\left(\frac{\partial}{\partial X}\right)^{-1}\right)w\label{QTEigen}
\end{eqnarray}

\subsection{Quantum transitions}
In general, there is a possibility of transition between the quantum states. Consider two
states, designate them $a$ and $b$. The probability of transition from state $a$ to state $b$ is
$P_{ab}=Tr(\opr\rho_a\opr\rho_b)= \int\rho_a(\vcc{q',q''})\rho_b(\vcc{q'',q'})d\vcc q'd\vcc q''$.
In terms of the Wigner formalism this can be rewritten as
\begin{equation}
P_{ab}=(2\pi)^{Nd}\int F^{W(a)}(\vcc{q,p})F^{W(b)}(\vcc{q,p})d\vcc qd\vcc p,\label{WFTransition}
\end{equation}
and recalling the connection of Wigner function with tomograms $w_1$ and $w$ (\ref{QTstoWF},
\ref{QTtoWF}) we easily get the following expressions for $P_{ab}$ in tomography approach:
\begin{eqnarray}
\int w^{a}(X,\vcc\mu,\vcc\nu)w^{b}(Y,-\vcc\mu,-\vcc\nu)
e^{i(X+Y)}\frac{dXdYd\vcc\mu d\vcc\nu}{(2\pi)^{Nd}}\label{QTTransition}
\end{eqnarray}

\subsection{Tomographic map in temperature-dependent processes}
The tomographic representation can be analogously introduced for the systems with temperature
$T\ne0$. In this case we consider 'imaginary time' $\beta=1/T$ (measuring $T$ in units of energy).
$\beta$ enters as a parameter in the density matrix, which is now defined by the equation
\begin{equation}
-\frac{\partial\rho(\vcc{q',q''},\beta)}{\partial\beta}=\opr H_{\vcc q'}\rho(\vcc{q',q''},\beta),\label{DMEvolT}
\end{equation}
where index $\vcc q'$ in $\opr H_{\vcc q'}$ shows that the Hamiltonian acts only on those variables.

Now the transition to the tomograms $w_1$ or $w$ is straightforward. We just use the same rules,
as in the derivation of evolution equation (\ref{QTEvol}) and eigenvalue equation
(\ref{QTEigen}). Then the evolution equation in imaginary time $\beta$ for $w$ is given by
\begin{eqnarray}
&&-\frac{\partial w}{\partial\beta}=\sum_{j=1}^{Nd}\left[\frac{1}{2m_j}\frac{\partial^2}
{\partial\nu_j^2}\left(\frac{\partial}{\partial X}\right)^{-2} -
\frac{1}{8m_j}\mu_j^2\frac{\partial^2}{\partial X^2}\right]w +\nonumber\\
&&ReV\left(\frac{i\vcc\nu}{2}\frac{\partial}{\partial X} - \frac{\partial}{\partial \vcc\mu}
\left(\frac{\partial}{\partial X}\right)^{-1}\right)w\nonumber\\
&&-\sum_{j=1}^{Nd}\frac{\mu_j}{2m_j}\frac{\partial w}{\partial\nu_j}=\nonumber\\
&&ImV\left(\frac i 2\vcc\nu\frac{\partial}{\partial X} - \frac{\partial}{\partial \vcc\mu}
\left(\frac{\partial}{\partial X}\right)^{-1}\right)w\label{QTEvolT}
\end{eqnarray}

Initial condition for Eq.(\ref{DMEvolT}) is $\rho(\vcc{q',q''},\beta=0)=\delta(\vcc{q'-q''})$.
It corresponds to constant Wigner function (see Eq.(\ref{DMtoWF})). Using Eq.(\ref{WFtoQT}) we
see that the tomogram $w$ for $\beta=0$ must have the delta-function form, equal zero everywhere,
besides the point $\vcc{\mu,\nu}=0$ and constant in $X$ direction in that point.

\subsection{Average values calculation}\label{Average}
Developing the 'center-of-mass' tomography formalism we must provide the rules of average values
calculation to complete the picture. Using the density matrix to describe the state of the system
we can obtain the average value of some operator $\opr A$ as
\begin{equation}
\avr A = Tr(\opr\rho\opr A), \label{AvrDM}
\end{equation}
where we choose $Tr(\opr\rho)=1$.

Here it is again convenient to begin with the Wigner-Moyal formulation of quantum mechanics. In
its framework to calculate the average value one deals with the {\it Weyl symbol}
$A^W(\vcc q,\vcc p)$ \cite{Weyl} of operator $A(\opr\vcc q,\opr\vcc p)$
(see \cite{TatarskiUFN,Lee1995} for review):
\begin{equation}
\avr A = \int A^W(\vcc q,\vcc p)W(\vcc q,\vcc p)d\vcc qd\vcc p, \label{AvrWF}
\end{equation}
where the Weyl symbol is given by
\begin{equation}
A^W(\vcc q,\vcc p) = \int Tr(A(\opr\vcc q,\opr\vcc p)e^{i\vcc{\xi\opr q}+i\vcc{\eta\opr p}})
e^{-i\vcc{\xi q}-i\vcc{\eta p}}\frac{d\vcc\xi d\vcc\eta}{(2\pi)^{2Nd}}\label{Weyl}
\end{equation}

Expression for the average values in one random variable tomography formulation is obtained
using the connection between $w$ and Wigner function (\ref{QTtoWF}):
\begin{eqnarray}
\avr A = \int e^{iX}w(X,\vcc\mu,\vcc\nu)A(\vcc\mu,\vcc\nu)dX d\vcc\mu d\vcc\nu,\label{AvrQT}\\
A(\vcc\mu,\vcc\nu)=\int A^W(\vcc q,\vcc p)e^{-i(\vcc\mu\vcc q + \vcc\nu\vcc p)}
\frac{d\vcc qd\vcc p}{(2\pi)^{2Nd}} \label{WeylF}
\end{eqnarray}

If considered operator depends on coordinates $\opr\vcc q$ or momenta $\opr\vcc p$ only, Weyl
symbols have the same form as corresponding operators in coordinate or momentum representation.
Operator $A(\opr\vcc q)$ is $A(\vcc x)$ in $\vcc x$-coordinate representation, then its Weyl symbol
$A^W(\vcc{q,p})$ is equal to $A(\vcc q)$. The same is valid for momenta-dependent operator:
$B(\opr\vcc p)$ is $B(\vcc y)$ in $\vcc y$-momentum representation, and $B^W(\vcc{q,p})=B(\vcc p)$.

Consider an operator $A(\opr\vcc q)$, depending on coordinates only. For momenta-dependent
operators all equations are the same, provided $\mu$ is replaced by $\nu$, and vice versa,
because the pairs $\vcc q,\vcc\mu$ and $\vcc p,\vcc\nu$ enter the equations connecting the
tomogram $w$ with Wigner function symmetrically. Integration over $\vcc\nu$ in Eq.(\ref{AvrQT})
for operator $A(\opr\vcc q)$ gives the delta-function $\delta(\vcc\nu)$. Then we have:
\begin{eqnarray}
\avr A = \int A^W(\vcc q)e^{-i(\vcc{\mu q}-X)}w(X,\vcc\mu,\vcc\nu=0)
\frac{dX d\vcc\mu d\vcc q}{(2\pi)^{Nd}} \label{AvrQTq}
\end{eqnarray}
It is often necessary to operate with the one-particle and one-dimension operators. Then, quite
generally, we can consider an operator $A(\opr q_1)$. Corresponding average value is given by
\begin{eqnarray}
\avr A = \int A^W(X)w(X,\mu_1=1,\tilde\vcc\mu=0,0)dX, \label{AvrQTq1}
\end{eqnarray}
where $\tilde\vcc\mu$ designates all $\mu_j$ except the specified $\mu_1$.

\section{Examples}\label{Examples}
In this section we introduce several examples of tomographic map for many-particles quantum
states. For simplicity, here we do not regard symmetry over particles exchange. Permutations
properties are considered in Sec.~\ref{Permutations}.

\subsection{Gaussian states}
Quite simple is the case of pure state and wave function of Gaussian form. This can be the ground
state of the system of independent oscillators, as well as coherent or squeezed states, or any
many-dimensional Gaussian wave packet. Such wave packet can be created due to parametric
excitation of multimode vacuum state of electromagnetic field \cite{FIANVol183}, {\it e.g.}, in
the framework of nonstationary Casimir effect \cite{DodReview}.

Consider the pure state with the wave function $\Psi(\vcc q) = \prod_{j=1}^{Nd}\psi_j(q_j)$, where
\begin{equation}
\psi_j(q) = (A_j/\pi)^{1/4}e^{-\frac{A_j}{2}(q-x_j)^2+iy_jq}\label{PsiGj}
\end{equation}
The only mathematical fact we need here is that the Fourier transform of a Gaussian
is Gaussian. Then, using Eq.(\ref{DMtoWF}) we immediately obtain the Wigner function as a product
of $W_j(q_j,p_j)$, where
\begin{equation}
W_j(q,p) = e^{-A_j(q-x_j)^2}e^{-B_j(p-y_j)^2}(A_jB_j)^{1/2}/\pi,\label{WFGj}
\end{equation}
and for states (\ref{PsiGj}) $B_j = 1/A_j$. For the set of parameters $\vcc x,\vcc y,\vcc A,
\vcc B$ applying Fourier transformation (\ref{WFtoQT}) to (\ref{WFGj}) we have:
\begin{equation}
w^{Gauss}(X,\vcc\mu,\vcc\nu) = e^{-(X-\vcc\mu\vcc x-\vcc\nu\vcc y)^2/C}/\sqrt{\pi C},\label{QTGj}
\end{equation}
where $C=\sum_{j=1}^{Nd}(\mu_j^2/A_j+\nu_j^2/B_j)$.

Thermal density matrix of independent oscillators is also Gaussian, but it is not a product
of wave functions, as the state is not pure. Still it is a product of density matrices of
individual oscillators (see, {\it e.g.} \cite{StatMechF}):
\begin{equation}
\rho_j(q,q') = \sqrt{\frac{2A_j(B_j-1)}{\pi}}e^{-A_j[B_j(q^2+q'^2)-2qq']},\label{DMTj}
\end{equation}
where $A_j = m\omega_j/(2sh(\omega_j\beta))$ and $B = ch(\omega_j\beta)$. Omitting the
straightforward calculations, we obtain the tomogram $w$ in the following form:
\begin{eqnarray}
&&w^{(\beta)}(X,\vcc\mu,\vcc\nu) = \frac{e^{-X^2/D}}{\sqrt{\pi D}},\label{QTT}\\
&&D = \sum_{j=1}^{Nd}\left(\frac{\mu_j^2}{2A_j(B_j-1)}+2\nu_j^2A_j(B_j+1)\right)\label{DQTT}
\end{eqnarray}

\subsection{Fock states}
The Fock states of light (the eigenstates in representation of photons number) correspond to
ground or excited states of multimode oscillator. The state is labeled by vector $\vcc n$ of
integer numbers and wave function has the form:
\begin{equation}
\Psi(\vcc q) = \prod^{Nd}_{j=1}\frac{e^{-q_j^2/2}H_{n_j}(q_j)}{\pi^{1/4}\sqrt{2^{n_j}n_j!}},\label{PsiF}
\end{equation}
where $H_m$ is the Hermit polynomial of $m$-th order. To obtain the tomogram for such state we
use the following facts. First, coherent state of an oscillator is described by the Gaussian wave
function and, correspondingly, by the Gaussian tomogram (see Eq.(\ref{QTGj})). Coherent state is
labeled by complex vector $\vcc\alpha=\vcc a+i\vcc b$ and parameters of Gaussian wave function in
coordinate representation (\ref{PsiGj}) are $x_j=\sqrt{2}a_j$ and $y_j=-\sqrt{2}b_j$. Second,
the wave function of coherent state (for simplicity, one dimension is considered here) is expanded in the basis of Fock states as
\begin{equation}
\ket\alpha = e^{-|\alpha|^2/2}\sum_{n=0}^{\infty}\frac{\alpha^n}{\sqrt{n!}}\ket n,\label{AlphaF}
\end{equation}
which is connected with the expression for generating function of Hermit polynomials:
\begin{equation}
e^{-\alpha^2+2\alpha q} = \sum_{n=0}^{\infty}\frac{\alpha^n}{n!}H_n(q)\label{AlphaHn}
\end{equation}

Expanding the tomogram of coherent state in Hermit polynomials and wave function of coherent
state in corresponding integral expression in wave functions of Fock states we have
\begin{eqnarray}
&&w^{\vcc n}(X,\vcc\mu,\vcc\nu) = \int\delta(X-\sum_{j=1}^{Nd}X_j)\times\nonumber\\
&&\prod_{j=1}^{Nd}\frac{H^2_{n_j}\left(\frac{X_j}{\sqrt{\mu^2_j+\nu^2_j}}\right)
e^{\frac{-X^2_j}{(\mu^2_j+\nu^2_j)}}}{2^{n_j}n_j!\sqrt{\pi(\mu^2_j+\nu^2_j)}}d\vcc X\label{QTF}
\end{eqnarray}

For example, for $N=2, d=1$ and states with $n_1,n_2$ equal to $0$ or $1$ (denoted $(n_1,n_2)$)
the tomograms $w(X,\mu_1,\mu_2,\nu_1,\nu_2)$ have the forms
\begin{eqnarray}
&&w^{(0,0)}_2 = \frac{exp\left[-X^2/C\right]}{\sqrt{\pi C}},\label{QTF00}\\
&&w^{(0,1)}_2 = \sqrt{\frac{C_2}{\pi C_1}}\frac{(2C_2X^2+C_1C_2+C_1^2)e^{-\frac{X^2}{C}}}
{C^{5/2}},\label{QTF01}\\
&&w^{(1,1)}_2=\frac{4C_1^2C_2^2e^{-\frac{X^2}{C}}}{\sqrt \pi C^{5/2}}\times\nonumber\\
&&\left(\frac{X^4}{C^2}+\frac{X^2}{C}\frac{C_1^2+C_2^2-4C_1C_2}{C_1C_2}+\frac 3 4\right),\label{QTF11}
\end{eqnarray}
where $C_1=\mu_1^2+\nu_1^2$, $C_2=\mu_2^2+\nu_2^2$ and $C=C_1+C_2$.

\section{Symmetry properties with respect to particles permutations}\label{Permutations}
Consideration of identical particles exchange imposes the restrictions concerning the possible
form of the state-describing functions. In this section we discuss the corresponding properties
of one-random-variable tomogarphic map (see \cite{QTExchange2002} for permutation symmetry
properties of the symplectic tomogram).

Further we use the following notations. A vector without index $\vcc a$ has $Nd$ components,
vector with index $\vcc a_j$ denotes the set of some values, corresponding to $j$-th particle,
and consists of $d$ components. A vector $\tilde\vcc a$ denotes the collection of all
components of $\vcc a$, except those that are specified in the same expression. For example,
$\tilde\vcc q$ in the expression $\psi(\vcc q_j,\tilde\vcc q)$ is the vector of all the coordinates,
except the coordinates of the $j$-th particle.

For particles obeying Fermi or Bose statistics, we have the following symmetry properties
concerning their permutations:
\begin{eqnarray}
\rho(\vcc q_j',\vcc q_i',\tilde\vcc{q'};\vcc q_i'',\vcc q_j'',\tilde\vcc{q''}) = \rho(\vcc q_i',
\vcc q_j',\tilde\vcc{q'};\vcc q_j'',\vcc q_i'',\tilde\vcc{q''})=\nonumber\\ \pm\rho(\vcc q_i',
\vcc q_j',\tilde\vcc{q'};\vcc q_i'',\vcc q_j'',\tilde\vcc{q''}),\label{DMFB}
\end{eqnarray}
where the upper sign ('$+$') is for Bose systems, and lower sign ('$-$') is for Fermi systems.
Note that 'entire' particles permutation (two particles exchange both $q$ and $q'$ variables)
corresponds to {\it sign conservation for both Fermi and Bose statistics}:
\begin{eqnarray}
\rho(\vcc q_j',\vcc q_i',\tilde\vcc{q'};\vcc q_j'',\vcc q_i'',\tilde\vcc{q''}) =
\rho(\vcc q_i',\vcc q_j',\tilde\vcc{q'};\vcc q_i'',\vcc q_j'',\tilde\vcc{q''})\label{DMFB2}
\end{eqnarray}

In the expressions for obtaining the Wigner function form density matrix (\ref{DMtoWF}) and
tomogram $w$ from Wigner function (\ref{WFtoQT}) we can exchange the
integration variables ($\vcc u_j \leftrightarrow \vcc u_i$, etc.), then we immediately have:
\begin{eqnarray}
W(\vcc q_j,\vcc q_i,\tilde\vcc q;\vcc p_j,\vcc p_i,\tilde\vcc p) =
W(\vcc q_i,\vcc q_j,\tilde\vcc q;\vcc p_i,\vcc p_j,\tilde\vcc p),\label{WFFB2}\\
w(X;\vcc\mu_j,\vcc\mu_i,\tilde\vcc\mu;\vcc\nu_j,\vcc\nu_i,\tilde\vcc\nu)=
w(X;\vcc\mu_i,\vcc\mu_j,\tilde\vcc\mu;\vcc\nu_i,\vcc\nu_j,\tilde\vcc\nu)\label{QTFB2}
\end{eqnarray}

We see that there is no distinction between Fermi and Bose statistics when the particles exchange
'entirely', {\it i.e.} $q$ and $q'$ in the density matrix, $q$ and $p$ in Wigner function or
$\mu,\nu$ in $w$ are permuted {\it simultaneously}. The distinction appears when not all the
variables, corresponding to the considered particles, are permuted. When we use the density
matrix, Fermi and Bose statistics differ only in the sign $\pm 1$, which appears after the
permutation of either $\vcc q_i',\vcc q_j'$ or $\vcc q_i'',\vcc q_j''$. For the Wigner function
and tomogram this difference is expressed in far more complicated manner, through the integral
transforms (see corresponding formulae for the symplectic tomography in \cite{QTExchange2002}).

First, regard the permutation of $\vcc q_i,\vcc q_j$ or $\vcc p_i,\vcc p_j$ for the Wigner
function. Again exchanging the integration variables in (\ref{DMtoWF}) we come to
\begin{equation}
W(\vcc q_j,\vcc q_i,\tilde\vcc q;\vcc p_i,\vcc p_j,\tilde\vcc p) =
W(\vcc q_i,\vcc q_j,\tilde\vcc q;\vcc p_j,\vcc p_i,\tilde\vcc p)\label{WFFBqp}
\end{equation}

The same considerations lead us to the similar expression for $w$:
\begin{equation}
w(X;\vcc\mu_j,\vcc\mu_i,\tilde\vcc\mu;\vcc\nu_i,\vcc\nu_j,\tilde\vcc\nu) =
w(X;\vcc\mu_i,\vcc\mu_j,\tilde\vcc\mu;\vcc\nu_j,\vcc\nu_i,\tilde\vcc\nu)\label{QTFBmunu}
\end{equation}

Then it is enough to develop the formulae for coordinate (Wigner function) or $\vcc\mu$ (tomogram)
permutations only. Corresponding integral expressions has the following form:
\begin{widetext}
\begin{eqnarray}
&&W(\vcc q_j,\vcc q_i,\tilde\vcc q;\vcc p_i,\vcc p_j,\tilde\vcc p) =
\int K^W(\vcc x_i,\vcc x_j,\vcc y_i,\vcc y_j,\vcc q_i,\vcc q_j,\vcc p_i,\vcc p_j)
W(\vcc x_i,\vcc x_j,\tilde\vcc q;\vcc y_i,\vcc y_j,\tilde\vcc p)
d\vcc x_id\vcc x_jd\vcc y_id\vcc y_j,\label{WFIntPerm}\\
&&w(X,\vcc\mu_j,\vcc\mu_i,\tilde\vcc\mu,\vcc\nu_i,\vcc\nu_j,\tilde\vcc\nu) =
\int K(X,\vcc\mu_i,\vcc\mu_j,\vcc\nu_i,\vcc\nu_j;Y,\vcc\xi_i,\vcc\xi_j,\vcc\eta_i,\vcc\eta_j)
w(Y,\vcc\xi_i,\vcc\xi_j,\tilde\vcc\mu,\vcc\eta_i,\vcc\eta_j,\tilde\vcc\nu)
dYd\vcc\xi_id\vcc\xi_jd\vcc\eta_id\vcc\eta_j,\label{QTIntPerm}
\end{eqnarray}
and kernels are given by
\begin{eqnarray}
&&K^W(\vcc x_i,\vcc x_j,\vcc y_i,\vcc y_j,\vcc q_i,\vcc q_j,\vcc p_i,\vcc p_j)=
\pm\left(\frac{4}{2\pi}\right)^d\delta(\vcc x_i+\vcc x_j-\vcc q_i-\vcc q_j)
\delta(\vcc y_i+\vcc y_j-\vcc p_i-\vcc p_j)
e^{i[(\vcc q_i-\vcc q_j)(\vcc y_i-\vcc y_j)+(\vcc x_i-\vcc x_j)(\vcc p_i-\vcc p_j)]}\label{WFKP}\\
&&K(X,\vcc\mu_i,\vcc\mu_j,\vcc\nu_i,\vcc\nu_j;Y,\vcc\xi_i,\vcc\xi_j,\vcc\eta_i,\vcc\eta_j)=\nonumber\\
&&\pm\int\frac{|k|^{2d}}{(2\pi)^{d+1}}\delta(\vcc\xi_i+\vcc\xi_j-\vcc\mu_i-\vcc\mu_j)
\delta(\vcc\eta_i+\vcc\eta_j-\vcc\nu_i-\vcc\nu_j)
e^{-i\{k(X-Y)-k^2/4[(\vcc\mu_i-\vcc\mu_j)(\vcc\eta_i-\vcc\eta_j)+
(\vcc\xi_i-\vcc\xi_j)(\vcc\nu_i-\vcc\nu_j)]\}}dk\label{QTKP}
\end{eqnarray}
\end{widetext}

\section{Conclusion}\label{Conclusion}
We studied in details the version of tomographic map of the density matrix and Wigner function
for which the quantum state of multimode system is associated with probability distribution
function. This function depends on one random variable $X$ and $2Nd$ real parameters (real
$Nd$-vectors $\vcc\mu$ and $\vcc\nu$) and it determines the quantum state completely. It means
that provided this probability distribution function is known one can reconstruct the Wigner
function of the system state and corresponding density operator. The random variable $X$ can be
interpreted as the system "center of mass" coordinate considered in specifically rotated and
scaled reference frame in the complete phase space of the system. Real parameters (vectors
$\vcc\mu$ and $\vcc\nu$) determine this rotated and scaled reference frame.

Information contained in the introduced tomogram ($w$) is the same as that contained in the
symplectic tomogram ($w_1$), which depends on larger number of variables. It corresponds to the
fact that the tomograms have high symmetry properties. By means of the symmetry operations one
can reconstruct the dependence of the function on larger number of variables starting from
initial function with smaller number of variables.

We have constructed the quantum evolution equations and energy level equations for the introduced
"center of mass" tomogram. Example of multimode oscillator and symmetry properties of the
tomogram for identical particles (fermions and bosons) were discussed in details.

\begin{acknowledgments}
We deeply appreciate the financial help from RFBR. A.A. is also grateful to 'Dynasty' foundation
and ICFPM for financial support.
\end{acknowledgments}

\appendix

\end{document}